\documentclass[aps,prx,twocolumn,showpacs,superscriptaddress]{revtex4-2}

\usepackage{graphicx}
\usepackage{dcolumn}
\usepackage{bm}
\usepackage[colorlinks=true, linkcolor=blue, citecolor=blue, urlcolor=blue]{hyperref}
\usepackage{lineno}

\begin{document}
\title{Fast Recovery of Niobium-based Superconducting Resonators after Laser Illumination}


\author{Chunzhen Li}
\affiliation{Department of Electrical Engineering, Yale University, New Haven, Connecticut 06520, USA}

\author{Yuntao Xu}
\affiliation{Department of Electrical Engineering, Yale University, New Haven, Connecticut 06520, USA}

\author{Yufeng Wu}
\affiliation{Department of Electrical Engineering, Yale University, New Haven, Connecticut 06520, USA}

\author{Manuel C. C. Pace}
\affiliation{Department of Electrical Engineering, Yale University, New Haven, Connecticut 06520, USA}

\author{Matthew D. LaHaye}
\affiliation{Air Force Research Laboratory, Information Directorate, Rome, New York 13441, USA}

\author{Michael Senatore}
\affiliation{Air Force Research Laboratory, Information Directorate, Rome, New York 13441, USA}

\author{Hong X. Tang}
\email{hong.tang@yale.edu}
\affiliation{Department of Electrical Engineering, Yale University, New Haven, Connecticut 06520, USA}

\begin{abstract}

Interfacing superconducting microwave resonators with optical systems enables sensitive photon detectors, quantum transducers, and related quantum technologies. Achieving high optical pulse repetition is crucial for maximizing the device throughput. However, light-induced deterioration, such as quasiparticle poisoning, pair-breaking-phonon generation, and elevated temperature, hinders the rapid recovery of superconducting circuits, limiting their ability to sustain high optical pulse repetition rates. Understanding these loss mechanisms and enabling fast circuit recovery are therefore critical. In this work, we investigate the impact of optical illumination on niobium nitride and niobium microwave resonators by immersing them in superfluid helium-4 and demonstrate a three-order-of-magnitude faster resonance recovery compared to vacuum. By analyzing transient resonance responses, we provide insights into light-induced dynamics in these superconductors, highlighting the advantages of niobium-based superconductors and superfluid helium for rapid circuit recovery in superconducting quantum systems integrated with optical fields.

\end{abstract}

\maketitle


\section{Introduction}

Superconducting microwave resonators integrated with photonic circuits are pivotal components exploited in microwave-to-optical data links \cite{elshaari2020hybrid,shen2024photonic}. Applying optical pulses at high repetition rates is essential for enhancing system performance and maximizing data throughput. For instance, in single-photon detectors \cite{hadfield2009single,sprengers2011waveguide,you2020superconducting}, high-speed optical excitation enhances photon count rate and timing resolution. Similarly, in microwave-to-optical transducers \cite{rueda2016efficient,fan2018superconducting,xu2021bidirectional,sahu2022quantum}, high optical pulse repetition rates enable faster quantum state transfer between superconducting circuits and optical communication channels, facilitating scalable quantum networks \cite{schoelkopf2008wiring,zhong2020proposal,jiang2007distributed,cirac1997quantum}. However, light-induced losses in superconductors, such as quasiparticle (QP) generation \cite{il2000picosecond,beck2011energy,barends2011minimizing,kardakova2013electron,PhysRevLett.133.060602}, pair-breaking-phonon generation \cite{kozorezov2000quasiparticle,patel2017phonon,iaia2022phonon} and temperature elevation, can all degrade the performance of superconducting microwave resonators and hinder their rapid recovery following laser illumination. 

QP generation arises from Cooper pair breaking when photons are directly absorbed by superconductors. The presence of excess QPs increases resistive losses, leading to decreased resonance quality factor ($Q$). Meanwhile, phonon generation from QP recombination and inelastic scattering of QPs extends the effective lifetime of QPs due to pair-breaking and contributes to heat accumulation, as the dissipation of excess thermal energy is constrained by the \textcolor{black}{thermal boundary resistance} \cite{schmidt1976thermal,sahling1981thermal,swartz1989thermal} at the superconductor-substrate interface. This limited heat transfer exacerbates local heating, further degrading superconducting circuit performance. These effects impose fundamental limitations on the achievable pulse repetition rates and overall system performance. However, distinguishing between different loss mechanisms is challenging, which complicates the development of effective mitigation strategies. Therefore, understanding light-induced dynamics, disentangling the interplay of various loss mechanisms, and mitigating light-induced loss in superconductors are crucial for advanced superconducting-photonic hybrid technologies and ensuring their viability for high-speed quantum information processing \cite{monroe2002quantum,wendin2017quantum,blais2020quantum}.

In this work, we investigate the transient response of niobium nitride (NbN) and niobium (Nb) microwave resonators to laser pulses when immersed in superfluid helium-4 ($^4\mathrm{He}$). By varying the laser pulse width, we demonstrate a three-order-of-magnitude faster resonance recovery for both resonators in superfluid $^4\mathrm{He}$ compared to vacuum. Since both NbN and Nb exhibit intrinsically short QP lifetimes on the nanosecond timescale \cite{men1997superconducting,ferrari2017hot,uzawa2020optical,haberkorn2024probing,kaplan1976quasiparticle,warburton1995quasiparticle,lobo2005photoinduced,vardulakis2007superconducting,barends2009photon}, we distinguish between QP dynamics and phonon-induced heating effects by analyzing their distinct time constants. In vacuum, rapid frequency shifts occurring on the nanosecond timescale at the laser pulse rise and fall edges are primarily attributed to QP dynamics, whereas slower resonant frequency shifts on the millisecond timescale during the laser pulse-on and -off periods indicate local heating and cooling effects related to phonon dynamics. In superfluid $^4\mathrm{He}$, the extended superconductor-helium interface greatly enhances thermal phonon escape rates from the superconductor, while superfluid $^4\mathrm{He}$’s exceptionally high thermal conductivity \cite{anderson1966considerations,brooks1977calculated,mongiovi2018non,minowa2022optical} enables efficient transport of thermal phonons to the environment. As a result, the resonators maintain stable resonant frequencies throughout laser illumination. These findings provide crucial insights into light-induced loss mechanisms in niobium-based superconducting resonators and demonstrate that superfluid $^4\mathrm{He}$ can effectively mitigate thermal effects, offering a promising strategy for enhancing the stability and performance of superconducting circuits in optically integrated quantum technologies.

\begin{figure}[t]
\centering
\includegraphics[width = 1\linewidth]{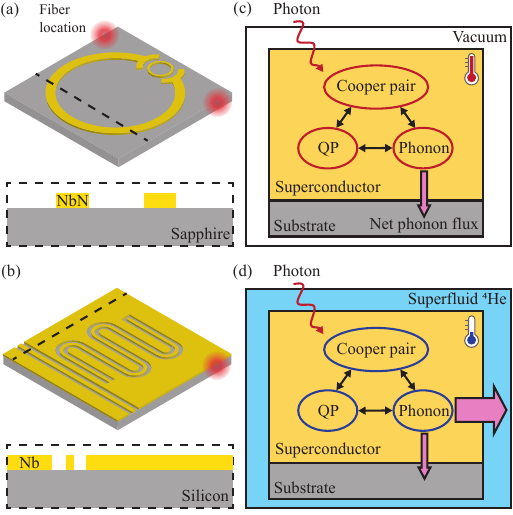}
\caption{\label{device} 
(a) Schematic drawing of the NbN resonator under study (not to scale). This resonator is linked to a coplanar waveguide (CPW, not shown) for RF coupling and is part of the on-chip electro-optics transducer device as introduced in reference \cite{xu2022light}. Laser light is coupled to the chip via a pair of \textcolor{black}{LN} grating couplers \textcolor{black}{near the chip edge} (not shown). The cross section along the dashed line is shown at the bottom. (b) Schematic drawing of the Nb resonator under study (not to scale). This resonator is a quarter-wave CPW resonator coupled to a transmon qubit (not shown). Laser light is delivered to the chip through \textcolor{black}{a single-mode} fiber \textcolor{black}{glued perpendicular to the side facet of the substrate}. The cross section along the dashed line is shown at the bottom. (c) and (d) Conceptual diagrams illustrating the underlying mechanisms when the superconducting devices are exposed to a laser pulse in vacuum and in superfluid $^4\mathrm{He}$, respectively. ``QP'' refers to quasiparticle.
}
\end{figure}

\section{Results}


\subsection{Sample design}

The NbN and Nb superconducting microwave resonators studied in this work are fabricated on separate chips and measured at different times. The NbN resonator is part of an on-chip electro-optic transducer device fabricated on a thin-film lithium niobate (TFLN)-NbN hybrid material system as introduced in reference \cite{xu2022light}. The device includes a \textcolor{black}{50-nm} thin-film NbN superconducting microwave resonator coupled to a pair of full-etch TFLN racetrack optical resonators. \textcolor{black}{The NbN resonator is measured with high radio frequency (RF) power. Its loaded quality factor $Q_\mathrm{tot}$ is 2.1\,k and intrinsic quality factor $Q_\mathrm{in}$ is 25.4\,k in vacuum.} The substrate for this device is sapphire, and laser light is sent to and collected from the device through a pair of \textcolor{black}{LN} grating couplers \cite{cheng2020grating} \textcolor{black}{near the chip edge} glued with angled lensed fibers \cite{mckenna2019alignment}, with a total \textcolor{black}{on-chip} transmission of -22.0 dB at 1560 nm at \textcolor{black}{millikelvin (mK)}. \textcolor{black}{This indicates that a significant portion of the light is dissipated near the coupling interface.} Its schematic drawing is shown in Fig.~\ref{device}(a). The Nb resonator is a quarter-wave coplanar waveguide resonator \textcolor{black}{with a thickness of 165 nm}, serving as a readout resonator coupled to a transmon qubit on a silicon substrate. Its schematic drawing is shown in Fig.~\ref{device}(b). To isolate the resonator's response from the qubit, the Nb resonator is \textcolor{black}{also} operated with high \textcolor{black}{RF} power to ensure it remains effectively decoupled. \textcolor{black}{Its $Q_\mathrm{tot}$ is 104.9\,k and $Q_\mathrm{in}$ is 213.0\,k in vacuum.} Since this chip lacks integrated optical waveguides, laser light is delivered \textcolor{black}{via a single-mode fiber glued perpendicular to the side facet of the silicon substrate}, fixed at 1550 nm. \textcolor{black}{Despite differences in coupling geometry, both devices experience non-local, scattered optical excitation. Since sapphire and silicon both exhibit low absorption at 1550-1560 nm \cite{merberg1993optical,soref2002silicon}, the incident light likely undergoes multiple reflections and scattering within the substrate, broadly illuminating the resonators and making the overall illumination conditions qualitatively similar.} Each device is \textcolor{black}{secured using adhesive and} mounted in a superfluid-tight copper cell equipped with RF and optical feed-throughs, \textcolor{black}{where the RF feed-throughs are sealed with indium and the optical ones use Stycast epoxy \cite{he2020micro-resonator,wasserman2022cryogenic}. The helium cell design and packaging details are similar to those in \cite{yang2020circuit,zhou2022single}.} Each cell is thermally anchored to the mixing chamber (MXC) of a dilution refrigerator. The MXC temperature is set at 150 mK using a PID-controlled heater, ensuring stable operating conditions for the measurements. \textcolor{black}{$^4\mathrm{He}$ is delivered shot-by-shot through a thin stainless steel capillary line that extends from room temperature down to the top of the helium cell, with thermal anchoring at each stage of the dilution refrigerator.}

\subsection{In Vacuum}

\begin{figure}[t]
\centering
\includegraphics[width = 1\linewidth]{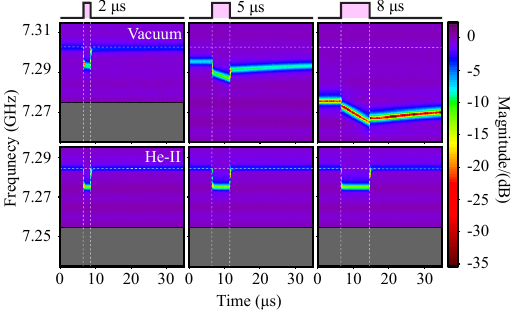}
\caption{\label{NbN} The time evolution of the NbN resonator's resonance over 35\,{\textmu}s. The upper three panels show the results when the device is vacuum, and the lower three panels show the results when the device is immersed in superfluid $^4\mathrm{He}$. The laser pulse width is varied from 2\,{\textmu}s, 5\,{\textmu}s to 8\,{\textmu}s. The laser pulse repetition period and peak power is fixed at 1 ms and 17 dBm. The original resonant frequencies in the absence of light in both cases are marked as the horizontal dashed lines in the plot. The vertical dashed lines mark the laser pulse rise and fall edges in different situations.
}
\end{figure}

We first examine the time evolution of the resonance for the NbN resonator with laser pulses in vacuum. As shown in Fig.~\ref{NbN}, the laser pulse configuration is depicted at the top, where the pulse width is varied from 2\,{\textmu}s to 5\,{\textmu}s and 8\,{\textmu}s. The laser pulse repetition period and peak power \textcolor{black}{at the fiber-chip interface} are fixed at 1 ms and 17 dBm, respectively. At the laser pulse rise edge, a rapid resonant frequency downshift is observed. \textcolor{black}{With a 2\,{\textmu}s-wide laser pulse, measured 1\,{\textmu}s after the pulse onset, the resonance $Q_\mathrm{tot}$ and $Q_\mathrm{in}$ decrease to 1.4\,k and 8.7\,k, respectively}. The resonant frequency continues to downshift as long as the laser pulse remains on. \textcolor{black}{The resonance is measured via $S_{11}$ reflection and is initially in the over-coupled ($Q_\mathrm{in} > Q_\mathrm{ex}$) regime. Since $Q_\mathrm{ex}$ remains stable, a decrease in $Q_\mathrm{in}$ due to light-induced losses brings the system closer to critical coupling, leading to a deeper resonance dip.} \textcolor{black}{With the 5(8)\,{\textmu}s-wide laser pulse, the resonance $Q_\mathrm{tot}$ and $Q_\mathrm{in}$ decrease to 1.3(1.1)\,k and 4.2(2.1)\,k, respectively.} The overall frequency downshift become more pronounced as the laser pulse width increases. At the laser pulse fall edge, a rapid resonant frequency upshift occurs, which shifts the same amount as the frequency downshift observed at the laser pulse rise edge. Following this rapid frequency upshift, both the resonant frequency and $Q$ gradually recover. When the laser pulse width is 2\,{\textmu}s, the resonance can fully recover before the next laser pulse arrives. However, as the laser pulse width increases, the cumulative heating shifts the resonance further away from its original state while the rapid frequency shifts at the laser pulse rise and fall edges become smaller. 

Possible underlying mechanisms of the observed phenomena in vacuum are depicted in Fig.~\ref{device}(c). 
\textcolor{black}{The} energy of an infrared photon at 1560 nm ($h\cdot192.3$ THz, where $h$ is the Planck constant) is two orders of magnitude higher than the superconducting energy gap ($2\Delta$) of NbN ($h\cdot1.2$ THz) \cite{uzawa2020optical} and Nb ($h\cdot1.0$ THz) \cite{thiemann2014niobium}. At the laser pulse rise edge, scattered photons are directly absorbed by the superconductor. These photons break Cooper pairs and generate QPs. The increased QP density increases the kinetic inductance of the superconductor, causing the decrease of the resonant frequency \cite{valenti2019interplay}. This Cooper-pair-photon interaction is an ultrafast process. For thin-film NbN \cite{il2000picosecond,beck2011energy} and Nb \cite{johnson1991photoresponse}, their photo-response times have been reported to be less than 1 ns. The excess energy of the photo-excited states is redistributed in the superconductor through scattering. This interaction quickly establishes a thermalized charged carrier distribution on the femtosecond timescale \cite{riffe2023excitation,takeda2024ultrafast}. Simultaneously, these energetic QPs recombine and inelastically scatter accompanied by pair-breaking phonon emission. The pair-breaking phonons down convert into low energies by breaking additional Cooper pairs or inelastically scattering, causing further decrease of the resonant frequency. In our measurement, the frequency transition time is mainly defined by the laser pulse rise time, which is estimated to be 70 ns. Consequently, we cannot separate the influence of QPs and pair-breaking phonons. In addition, increasing the laser pulse width raises the superconductor's temperature, leading to a reduction of Cooper pair density and a corresponding decrease in light-generated QP density \cite{de2011number,budoyo2016effects}. As a result, a smaller frequency downshift is observed at the rise edge of the laser pulse as the laser pulse width increases.

During the laser pulse-on period, thermal phonons down-converted from pair-breaking phonons should be the dominant loss source. The escape of thermal phonons to the substrate is limited by the \textcolor{black}{thermal boundary resistance} at the superconductor-substrate interface, leading to gradual heat accumulation in the superconductor during the laser pulse-on period. This results in a slow decrease of the resonant frequency and a degradation of $Q$. 

At the laser pulse fall edge, the circuit recovery undergoes two phases. In the first phase, QPs rapidly recombine while pair-breaking phonons quickly down-convert. NbN’s and Nb's QP lifetimes are typically on the nanosecond timescale, allowing rapid QP recombination. The fast resonant frequency upshift in the first phase suggests that the pair-breaking phonon down-conversion rate is high as well. In the second phase, the slow phonon escape rate limited by the \textcolor{black}{thermal boundary resistance} at the superconductor-substrate interface impedes efficient heat dissipation and therefore prevent the resonant frequency and $Q$ from rapidly recovering to its no-laser state following laser illumination.

We conduct a similar measurement with the Nb resonator in vacuum. \textcolor{black}{The Nb resonator follows a hanger-type design and is characterized through $S_{21}$ transmission measurements.} The results are shown in the upper three panels of Fig.~\ref{Nb}. The laser pulse configuration remains identical to that used for the NbN resonator, except that the laser pulse peak power \textcolor{black}{at the fiber-chip interface} is reduced to 13 dBm. This reduction is necessary to maintain the resonance visibility, \textcolor{black}{as the $Q_\mathrm{in}$ of the Nb resonator is significantly lowered by light-induced losses, pushing the system into a strongly under-coupled ($Q_\mathrm{in} < Q_\mathrm{ex}$) regime during the laser pulse-on period, which results in a very shallow resonance dip with $Q_\mathrm{tot}=$ 7.6\,k in vacuum and 10.1k in superfluid $^4\mathrm{He}$.} We observe that the frequency variation of the Nb resonator is significantly smaller than that of the NbN resonator. This is because NbN has a notably higher kinetic inductance compared to Nb \cite{xu2019frequency,niepce2019high,frasca2023nbn}. As a result, the NbN resonator is more sensitive to the Cooper pair density change and shows a larger resonant frequency shift when light illuminates it. During the laser pulse-on and -off periods, phenomena similar to those observed in the NbN resonator can also be seen. However, there is an additional and unexpected feature: a resonant frequency upshift at the laser pulse rise edge and a resonant frequency downshift at the laser pulse fall edge. This suggests that mechanisms beyond QP and pair-breaking phonon dynamics are involved. Although the exact underlying mechanisms remain uncertain, \textcolor{black}{a well-established theory \cite{von1977saturation,strom1978low,phillips1987two} attributes this behavior to the coupling between a bath of two-level systems (TLS) and the external field via their electric-dipole moment. This model derives the temperature dependence of the sample's permittivity due to TLS, predicting an upward shift in the resonant frequency with increasing temperature at mK temperatures. Experimental studies across various platforms \cite{gao2008experimental,lindstrom2009properties,Fischer_Development_2023} have confirmed this TLS-induced temperature-dependent frequency shift. In our case, the observed phenomenon could arise from TLS located in the native oxide layer on the surface of the silicon substrate \cite{bruno2015reducing} or the Nb film \cite{murthy2022developing}. At the laser pulse rise edge, the fast frequency shift reflects a combination between TLS- and QP-induced effects. At lower temperatures, the TLS-induced positive shift typically dominates, while at higher temperatures, the QP-induced negative shift becomes more significant \cite{wang2013photon,poorgholam2024engineering,foshat2025characterizing}. The frequency upshift at the laser pulse rise edge suggests TLS-related effects dominate. However, sustained laser illumination during the laser pulse-on period raises the device temperature further, increasing QP density. This shifts the dominant mechanism to QP-related effects, causing a net frequency downshift. Related studies have also examined light-induced dynamics in Nb-based resonators \cite{savinov2016giant,kalhor2021active}, providing further context for our observations.}

\subsection{In Superfluid $^4\mathrm{He}$} 

After performing measurements in vacuum, we immerse both resonators in superfluid $^4\mathrm{He}$. The results for the NbN resonator are shown in the lower three panels of Fig.~\ref{NbN}. \textcolor{black}{Compared to that in vacuum, the resonant frequency in the absence of light decreases by approximately 19.04 MHz when the device is immersed in superfluid $^4\mathrm{He}$. This frequency shift is primarily due to the change in the dielectric environment above the chip. The $Q_\mathrm{tot}$ and $Q_\mathrm{in}$ are 2.2\,k and 16.5\,k, respectively. In addition,} the fast resonant frequency shifts at the laser pulse rise and fall edges remain present. However, the slow resonant frequency shifts and $Q$ changes observed during the laser pulse-on and -off periods in vacuum disappear. When the laser pulse is turned on, the resonance remains stable throughout. \textcolor{black}{The $Q_\mathrm{tot}$ and $Q_\mathrm{in}$ drop to 1.7\,k and 8.7\,k, respectively, which is comparable to that in vacuum with a 2\,{\textmu}s laser pulse.} Immediately after the laser pulse is turned off, both the resonant frequency and $Q$ quickly return to their original no-light values, with the frequency shift matching the magnitude observed at the laser pulse rise edge. Notably, the rapid frequency shifts at the laser pulse rise and fall edges, as well as the stable resonance during the pulse-on period, are unaffected by the laser pulse width. Additionally, the resonances can completely recover immediately after the laser pulse is turned off, regardless of its width.

Possible underlying mechanisms of the observed phenomena in superfluid $^4\mathrm{He}$ are illustrated in Fig.~\ref{device}(d). At the laser pulse rise edge, the fast resonant frequency shift indicates that superfluid $^4\mathrm{He}$ cannot help eliminate light-induced QP generation and pair-breaking phonon generation. The consistent fast resonant frequency shifts with different laser pulse widths suggest that superfluid $^4\mathrm{He}$ can efficiently thermalize the system to the base temperature \cite{lane2020integrating,lucas2023quantum}. 

\begin{figure}[t]
\centering
\includegraphics[width = 1\linewidth]{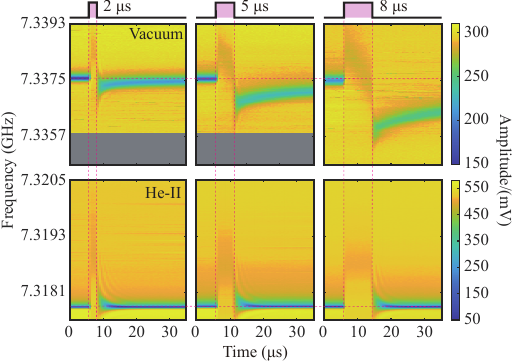}
\caption{\label{Nb} The time evolution of the Nb resonator's resonance over 35\,{\textmu}s. The upper three panels show the results when the device is vacuum, and the lower three panels show the results when the device is immersed in superfluid $^4\mathrm{He}$. The laser pulse width is varied from 2\,{\textmu}s, 5\,{\textmu}s to 8\,{\textmu}s. The laser pulse repetition period and peak power is fixed at 1 ms and 13 dBm. The original resonant frequencies in the absence of light in both cases are marked as the horizontal dashed lines in the plot. The vertical dashed lines mark the laser pulse rise and fall edges in different situations. Additionally, the color bars are rescaled to enhance the visibility of the resonance. 
}
\end{figure}

During the laser pulse-on period, immersing the device in superfluid $^4\mathrm{He}$ significantly enhances the superconductor-environment interface area \textcolor{black}{thanks to its zero-viscosity property \cite{kapitza1938viscosity,leggett1999superfluidity,gomez2014shapes}, reducing the total thermal resistance} and enabling efficient thermal phonon transport across the interface \cite{lucas2023quantum}. Thanks to the exceptionally high thermal conductivity of superfluid $^4\mathrm{He}$, the escaped thermal phonons rapidly dissipate, ensuring efficient heat transfer and effective thermalization of the device. Superfluid $^4\mathrm{He}$ suppresses heating caused by the accumulation of thermal phonons, thereby reducing the temperature rise in the superconductor and stabilizing the resonance. 

At the laser pulse fall edge, QPs quickly recombine and pair-breaking phonons rapidly down-convert as the case in vacuum. At the same time, thermal phonons also escape into and dissipate through superfluid $^4\mathrm{He}$, efficiently thermalizing the system to the base temperature. As a result, both the resonant frequency and $Q$ recover swiftly to their no-laser state. A closer inspection of the lower three panels in Fig.~\ref{NbN} reveals that the resonance fully recovers approximately 1\,{\textmu}s after the laser pulse fall edge with all the three laser configurations. In contrast, in vacuum, when the laser pulse width is set to 5\,{\textmu}s and 8\,{\textmu}s, the resonance fails to fully recover within the 1-ms repetition period. This highlights that the NbN resonator's recovery rate after laser illumination in superfluid $^4\mathrm{He}$ is three orders of magnitude faster than in vacuum, underscoring the exceptional heat dissipation capabilities of superfluid $^4\mathrm{He}$.

When the Nb resonator is immersed in superfluid $^4\mathrm{He}$, \textcolor{black}{the resonant frequency in the absence of light decreases by approximately 19.76 MHz. The $Q_\mathrm{tot}$ is 79.7\,k and the resonance enters a more over-coupled regime compared to vacuum, primarily due to a reduction in $Q_\mathrm{ex}$.} As shown in the lower three panels of Fig.~\ref{Nb}, we observe phenomena similar to those seen in the NbN resonator across all the three laser pulse widths. One key difference is that the positive resonant frequency shift \textcolor{black}{at the laser pulse rise edge} persists, as it does in vacuum. Interestingly, the magnitude of the resonant frequency shift at the laser pulse rise edge, approximately 1 MHz, is comparable in both vacuum and superfluid $^4\mathrm{He}$. This suggests that the mechanism driving the positive frequency shift is not significantly affected by the external environment. If the positive frequency shift is attributed to the coupling of the resonator to a bath of TLS, it implies that the TLS bath could arise from defects inside the materials and at the material interface \cite{pappas2011two,faoro2012internal,muller2019towards}, or trapped QPs \cite{de2020two}, rather than defects at the device surface.

Another notable difference is the clear observation of ripples at the laser pulse fall edge and a longer recovery time afterwards for the Nb resonator. These ripples arise from interference between microwave photons at the original resonant frequency ($f_o$) and those at the shifted resonant frequency ($f_s$) \cite{xu2022light}. \textcolor{black}{The resonant frequency shifts by 0.94 MHz, corresponding to a ripple period of 1.06\,{\textmu}s.} The mechanism behind this interference is as follows: when the resonator linewidth is small, as is the case for this Nb resonator, the cavity photon lifetime is relatively long. This leads to intracavity frequency conversion of the microwave photons before exiting the cavity. Considering a case where the system is being pumped at $f_s$. At the laser pulse fall edge, the frequency of the microwave photons inside the cavity rapidly change from $f_s$ to $f_o$. Then the pumping microwave photon at $f_s$ and the microwave photon at $f_o$ leaked from the cavity coexist in the output line, leading to interference. The intensity of this interference diminishes as the microwave photons at $f_o$ escape from the cavity. Therefore, the occurrence of the ripples means that the actual frequency transition at the laser pulse fall edge happens faster than the cavity photon lifetime, and it is obscured by the cavity ring-down process. The full recovery of the Nb resonator's resonance is about 10\,{\textmu}s. Although the circuit recovery is constrained by the resonance linewidth, the recovery rate of the Nb resonator in superfluid $^4\mathrm{He}$ is still three orders of magnitude faster than in vacuum.

\section{Discussion and conclusion}

In the measurements shown in Fig.~\ref{NbN} and Fig.~\ref{Nb}, we compare the photo-response of both types of resonators in vacuum and superfluid $^4\mathrm{He}$ as the laser pulse width increases. The results in superfluid $^4\mathrm{He}$ demonstrate its exceptional power-handling capabilities, efficiently managing light-induced heat accumulation and maintaining resonance stability. A natural question arises: what happens when the laser pulse repetition rate or peak power is varied? Increasing the laser pulse repetition rate effectively mimics increasing the laser pulse width, as both changes raise the duty cycle of the laser pulses. However, increasing the laser pulse peak power is different because it raises the energy delivered per laser pulse without altering the duty cycle. Higher peak power would lead to more light-induced QP generation, resulting in larger fast frequency shifts at the pulse rise and fall edges. The increased QP population would also drive stronger scatterings and interactions, pushing the system further away from equilibrium. When the laser peak power exceeds a certain threshold, the ability of superfluid $^4\mathrm{He}$ to dissipate heat may be overwhelmed. Beyond that point, the resonance during the laser pulse-on and -off period could become unstable even when the device is immersed in superfluid $^4\mathrm{He}$. \textcolor{black}{Beyond the result shown in Fig.~\ref{NbN}, we further increase the laser peak power at the fiber-chip interface to 22 dBm. Notably, no degradation in the cooling performance of superfluid $^4\mathrm{He}$ is observed, suggesting that power-handling capacity of the system remains well below saturation and the threshold is significantly higher than 22 dBm.} This robust performance highlights the efficiency of superfluid $^4\mathrm{He}$ in heat management. Despite not encountering any limitations in power handling with superfluid $^4\mathrm{He}$, we refrain from further increasing the laser pulse width due to the limited cooling power of the MXC. Thanks to the high critical temperatures ($T_c$) of NbN and Nb, future experiments could anchor the devices to the 1K plate of a dilution refrigerator. This would offer greater cooling power while retain the benefits of superfluid $^4\mathrm{He}$, paving the way for studies at even higher laser powers or extended pulse duty cycles.

To summarize, this work investigates the light-induced dynamics of NbN and Nb superconducting microwave resonators in both vacuum and superfluid $^4\mathrm{He}$. We study the time evolution of their resonant frequency and $Q$ under laser illumination, analyzing the possible underlying mechanisms of light-matter interactions in superconducting systems. The unique properties of superfluid $^4\mathrm{He}$ enable both types of resonators to demonstrate rapid recovery following laser illumination, with a rate three orders of magnitude faster than in vacuum. These findings deepen our understanding of the photo-response mechanisms in superconducting microwave resonators, providing insights into the interplay between photons, Cooper pairs, QPs and phonons under non-equilibrium conditions. The results underscore the potential of niobium-based superconducting materials, in conjunction with superfluid $^4\mathrm{He}$, for applications in high-speed superconducting-photonic hybrid quantum devices.

\section{Methods}

The measurement setup used to characterize the microwave resonators under different light illumination conditions consists of three sections: optical, RF and cryogenic. On the optical side, a tunable continuous-wave laser, set to either 1550 nm or 1560 nm, serves as the light source. Optical pulses are generated using an 80 MHz acousto-optic modulator (AOM), allowing precise control over pulse width and repetition rate. The modulated pulses are then delivered to the superconducting resonators via a fiber-coupled optical system. On the RF side, microwave signals near the resonant frequencies are generated using single-sideband modulation. This is achieved by feeding a low-frequency reference signal from a lock-in amplifier (LIA) into an IQ mixer, which upconverts it to the desired microwave frequency. The microwave response is measured at the output port, where the transmitted or reflected signal is amplified and detected. The output is demodulated by a mixer and then processed by the LIA, which extracts amplitude and phase information. The AOM and LIA are synchronized to enable precise time-resolved measurements of the resonance dynamics in response to laser pulses. Inside the dilution refrigerator, the input microwave signals are routed through attenuators and filters to ensure a well-defined, low-noise excitation before reaching the device. For the NbN resonator, the microwave output line consists of a traveling wave parametric amplifier followed by a high-electro-mobility transistor (HEMT) amplifier to provide high signal-to-noise ratio. For the Nb resonator, the output chain consists solely of a HEMT amplifier to amplify the device’s output signal. $^4\mathrm{He}$ is introduced through a thin stainless steel capillary line via a stainless tube on top of the cell.

\section{Acknowledgment}
This work was supported by the DOE Office of Science, the National Quantum Information Science Research Center, Codesign Center for Quantum Advantage (C2QA), Contract No. DESC0012704. HXT and YFW acknowledge funding support from the Air Force Office of Scientific Research (AFOSR No. MURI FA9550-23-1-0688). The JTWPAs used in this experiment are provided by IARPA and MIT Lincoln Laboratory. The authors thank Dafei Jin, Xu Han, and Xinhao Li for discussion on the implementation of the helium cells. The authors thank Dr. Yong Sun, Dr. Lauren McCabe, Mr. Kelly Woods, and Dr. Michael Rooks for assistance in device fabrication. ARFL discloses that this work is categorized as Approved for Public Release; Distribution Unlimited: PA\# AFRL-2025-1288. Any opinions, findings, and conclusions or recommendations expressed in this article are those of the authors and do not necessarily reflect the views of the Air Force Research Laboratory (AFRL). 

\section{Data availability}
The data that support the findings of this study are available from the corresponding author upon reasonable request.

\section{Code availability}
All relevant computer codes supporting this study are available from the corresponding author upon reasonable request.

\section{Author contributions}
H.X.T., C.L. and Y.X. conceived the experiment. C.L. packaged the Nb resonator. Y.X. fabricated and packaged the NbN resonator. C.L. and Y.X. performed the experiment. C.L., Y.X. and M.S. analyzed the data. Y.W., M.C.C.P. and M.D. helped with the project. C.L., M.S. and H.X.T. wrote the manuscript, and all authors contributed to the manuscript. All authors discussed the results and reviewed the manuscript. H.X.T. supervised the work.

\section{Competing interests}
The authors declare no competing interests.



\providecommand{\noopsort}[1]{}\providecommand{\singleletter}[1]{#1}%

\end{document}